\documentclass[aps,twocolumn,superscriptaddress]{revtex4}
\usepackage{graphicx}

\begin{document}
\title{Influence of fluctuations in actin structure on myosin V step size}
\author{Andrej Vilfan\\J. Stefan Institute, Jamova 39, 1000
  Ljubljana, Slovenia}
\date{1.5.2005}
\begin{abstract}
  We study the influence of disorder in the helical actin structure on the
  myosin V step size, predicted from the elastic lever arm model.  We show that
  fluctuations of $\pm 5$ degrees per actin subunit, as proposed by Egelman et
  al., significantly alter the distribution of step sizes and improve the
  agreement with experimental data.
\end{abstract}
\maketitle

\section*{Introduction}

Myosin V is a motor protein from the myosin superfamily, involved in various
intracellular transport processes \cite{Reck-Peterson.Mercer2000}. It is a
processive motor \cite{Mehta.Cheney1999}, which means that a single protein
molecule can transport cargoes along actin filaments over distances of several
micrometers.  It is dimeric, consisting of two identical heads, each attached
to a lever arm, joined with each other through a  tail domain.  It
achieves processivity by alternately binding its two heads to an actin filament
and thus walking in a hand-over-hand fashion \cite{Vale2003b}.  Its step size
is roughly determined by the periodicity of the actin filament, which is about
$36\,{\rm nm}$.

In comparison with muscle myosin (myosin II), two major adaptations are found
in myosin V: a longer lever-arm, measuring around $26\,{\rm nm}$
\cite{Burgess.Trinick2002} and a slower release of ADP.  The duty cycle of each
head is otherwise similar \cite{De_La_Cruz.Ostap2004}: the head binds to actin
in the ADP.Pi state, first releases Pi and performs a large conformational
change (power stroke), then releases ADP and performs a smaller conformational
change, and finally binds a new ATP molecule and detaches from actin.

In a recent article \cite{Vilfan2005}, we have developed a model for myosin V,
based on the elasticity of the lever arm.  It assumes that the lever arm is
stiffly anchored in the head, but the direction depends on the chemical state
of that head.  The distal ends of the lever arms are connected together through
a flexible hinge, which represents the only means of ``communication'' between
the heads.  By calculating the bending energies in all relevant dimer states
and its influence on transition rates, we have shown that the elastic lever-arm
model explains the coordinated hand-over-hand motility by showing that the lead
head is not likely to bind to actin before the trail head undergoes the major
power stroke and that it cannot commit its power stroke before the trail head
unbinds.  The model also quantitatively reproduces the measured force-velocity
relations and shows how the run length (the average distance a motor runs
processively before it dissociates from actin) could be used to determine some
kinetic rates.  Although the original model reproduces the step size
corresponding to about one half-turn of the actin helix (13 actin subunits),
there is a small but significant deviation between the predicted distribution
of step sizes (mainly on the 13th subunit, with a side peak on the 11th) and
the statistics obtained from electron microscopy (EM) studies
\cite{Walker.Knight2000}, which show the main peak on the 13th subunit and two
significant side peaks on the 11th and the 15th subunit, whereby the 15th is
stronger than the 11th.

\begin{figure}
  \begin{center}
    \includegraphics[width=8cm]{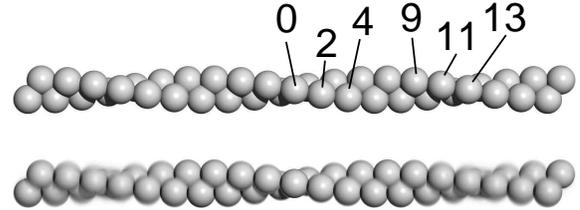}
  \end{center}
  \caption{The regular 13/6 helical model of an actin filament (top), and the
    28/13 helix with $\pm 5^\circ$ fluctuations per subunit (bottom).  The
    subunits are shown blurred, with a relative intensity corresponding to
    their statistical weight.  The orientation of the central subunit (denoted
    as ``0'') is always kept fixed.}
  \label{fig:twoactins}
\end{figure}

The purpose of the current paper is to extend the present model by including a
more precise description of the actin structure, which could potentially
explain the current deviation.  The central modification will be to take into
account torsional fluctuations in the actin structure.  So far we have used the
assumption that the actin monomers are arranged on a helix with a periodicity
of 13 actin subunits (also called a 13/6 helix, because the rotation of each
next subunit is $\frac{6}{13} \times 360^\circ$.  However, it has been known
for some time that the actin filaments have a variable, fluctuating twist
\cite{Egelman.DeRosier1982}.  The nature and the dynamics of these fluctuations
are not yet entirely understood, but it seems that the azimuthal orientation of
each subunit can fluctuate about $\pm 5^\circ$ with respect to its neighbors
\cite{Egelman.DeRosier1992}.  There is some evidence that the dynamics of these
fluctuations is rather slow \cite{Egelman1997,Orlova.Egelman2000}, with a
characteristic time of the order of seconds. This would imply that the twist
has discrete states with a considerable energy barrier between them.  The
average twist per subunit was found to be about $167^\circ$, slightly more than
in the 13/6 helix, where it would be $166.15^\circ$ (Fig.~\ref{fig:twoactins}).

\begin{figure*}
  \begin{center}
    \includegraphics[width=14cm]{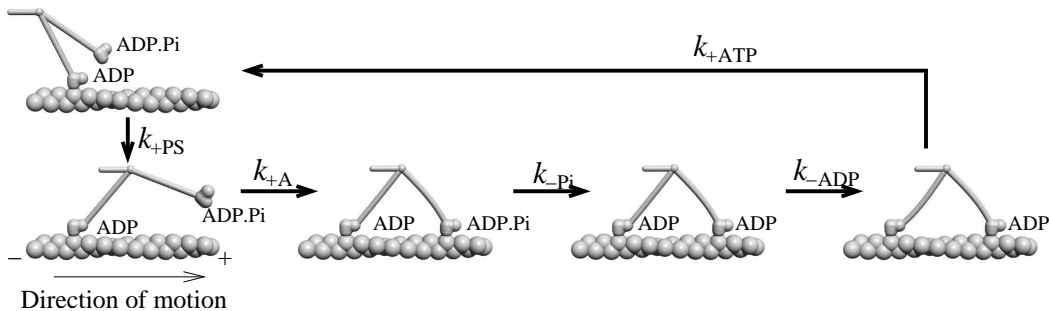}
  \end{center}
  \caption{The most likely duty cycle of the dimeric myosin V.  A head
    in the ADP state undergoes the power-stroke, then the second head
    containing ADP.Pi binds in the lead position, it releases Pi, then the rear
    head releases ADP (time limiting), binds a new ATP molecule and detaches
    from actin and the cycle repeats with exchanged roles of the two heads,
    however with the dimer having proceeded about one actin period further.}
  \label{fig:dutycycle}
\end{figure*}

Based on many experimental and theoretical studies
\cite{Rief.Spudich2000,Vale2003b,Rosenfeld.Sweeney2004,Vilfan2005}, the
following picture of the working cycle of dimeric myosin V has now found broad
consensus (Fig.~\ref{fig:dutycycle}).  The major part of the dimeric cycle is
spent in the state with both heads binding ADP.  The trail head is in the
post-powerstroke and the lead head likely in the pre-powerstroke state
\cite{Burgess.Trinick2002,Snyder.Selvin2005,Vilfan2005}, although the latter
has still been controversial recently (cf.~ref. \cite{Snyder.Selvin2004}).
Then the trail head releases ADP, binds a new ATP molecule and unbinds from
actin.  With the trail head released, the lead head is now free to undergo a
power-stroke, whereupon the former trail head binds in the lead position (one
step ahead in the direction of motion) and releases Pi. This brings the dimer
to the original state, however one step further toward the actin plus end and
with one ATP molecule hydrolyzed.

\begin{table}
\caption{Model parameters}
  \begin{tabular}{lll}
\hline
Lever arm length & $L$ & $26\,{\rm nm}$ for 6IQ (WT)\\
Lever arm  start (radial)   & $R$ & $8\,{\rm nm}$\\
Lever arm  start (longitudinal)   & $\delta_{\rm ADP.Pi}$ & $0\,{\rm nm}$\\
   & $\delta_{\rm ADP}$ & $3.5\,{\rm nm}$\\
Angle ADP.Pi & $\phi_{\rm ADP.Pi}$ & $115^{\circ}$\\
Angle ADP & $\phi_{\rm ADP}$ & $50^{\circ}$\\
Bending modulus & $EI$ & $1500\,\rm pN\,nm^2$\\
Average helix twist per subunit & $\theta^0$ & $167.14^\circ$\\
Twist fluctuations & $\theta'$ & $5^\circ$\\
Helix rise per subunit & $a$ & $2.75\,{\rm nm}$\\
\hline
\end{tabular}
\label{tab:parameters}
\end{table}

\begin{figure}
  \begin{center}
    \includegraphics[width=8cm]{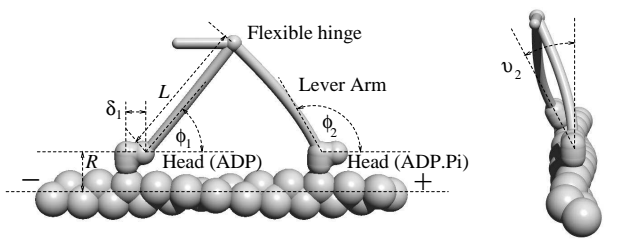}
  \end{center}
  \caption{The geometry of a dimeric myosin V molecule, after the lead head has bound to a site $13$ subunits in front of the trail head.}
  \label{fig:geometry}
\end{figure}

\section*{Results}

To calculate the step size, which is the main objective of this article, the
relevant transition is the binding of the lead head (in the ADP.Pi state) while
the trail head is in the forward-leaning ADP conformation.  Let us denote the
binding rate to the site $i$ (counted from the trail head, which is bound to
site $0$) with $k_{+A(i)}$ and the rate of the reverse (unbinding) process as
$k_{-A(i)}$.  From the detailed balance we know that the ratio of these rates
has to obey
\begin{equation}
  \label{eq:det_bal}
  \frac{k_{+A(i)}}{k_{-A(i)}}=  \frac{k^0_{+A}}{k^0_{-A}} \exp\left[ -
  \frac{\Delta U + F \Delta x}{k_B T} \right]
\end{equation}
where $\Delta U$ denotes the increase in elastic energy upon binding and
$F\Delta x$ denotes the work done against the applied load.  $k^0_{+A}$ and
$k^0_{-A}$ represent the binding and unbinding rates of a head that binds
without any elastic distortions in the lever arm.  Because the persistence
length of the lever arms is significantly longer than their actual length, we
neglect thermal fluctuations and the corresponding entropic contributions to
the total free energy.  In the following we will also concentrate on the case
of an unloaded motor ($F=0$).  In this case the bending energy in the initial
state is $0$ and $\Delta U$ is equal to the elastic energy in the double-bound
state.  All model parameters are summarized in Table~\ref{tab:parameters} and
Fig.~\ref{fig:geometry}.  The geometric parameters were obtained from published
EM studies \cite{Walker.Knight2000,Burgess.Trinick2002}.  The bending modulus of
the lever arm ($EI$) was estimated \cite{Vilfan2005} based on a lower boundary
provided by the fact that a myosin V molecule is able to perform regular steps
against loads of at least $1.8\,\rm pN$ \cite{mehta99}.

The elastic energy of a certain dimer conformation is calculated in the
following way (which is described in more detail in Ref.\cite{Vilfan2005}).  A
lever arm starting point is positioned at 
\begin{equation}
  \label{eq:16}
  \mathbf{x}^0=\left( \begin{array}{c}
      i a+\delta\\
       R \sin(\theta_i) \\
      R \cos(\theta_i)
    \end{array}
  \right)
\end{equation}
and its initial tangent is
\begin{equation}
  \label{eq:17}
  \hat{t}^0=\left( \begin{array}{c}
      \cos(\phi) \\
      \sin(\phi) \sin(\theta_i) \\
      \sin(\phi) \cos(\theta_i)
    \end{array}
  \right) \;.
\end{equation}
Other conditions state that the length of the lever arm from the starting point
to the joint has to be $L$ and that the endpoints of both lever arms coincide.
The bending energy is then given as
\begin{equation}
  \label{eq:energy}
  U=\sum_{j=1}^{2} \int_0^L \frac {EI}{2} \left( \frac{d \hat{t}_j}{ds}
  \right)^2 ds
\end{equation}
where $EI$ denotes the bending modulus and $\frac{d \hat{t}}{ds}$ the local
curvature of a lever arm.  The summation index $j$ runs over both lever arms.
The shape of the dimeric molecule is determined numerically in a way that
minimizes $U$.

There are two possibilities concerning the reversibility of the binding
process.  One possibility is that the lead head binds to a site and does not
detach before the next power stroke. In this case, the probability that it
binds to the site $i$ is
\begin{equation}
  \label{eq:bprob}
  P_i=\frac{k_{+A(i)}}{\sum_j k_{+A(j)}}\; .
\end{equation}
If, however, the lead head in the ADP.Pi (weakly bound) state is allowed to
detach and re-attach several times, the probability to find it on the site $i$
is given by the equilibrium distribution, 
\begin{equation}
  \label{eq:bprob1}
  P_i=\frac{k_{+A(i)}/{k_{-A(i)}} }{\sum_j k_{+A(j)}/k_{-A(j)} }\; .
\end{equation}

So far, no experimental data are available about the strain dependence of the
detachment rate, $k_{-A}$.  However, as with most binding processes one can
expect that the activation point for the binding transition is closer to the
bound state, and therefore the strain-dependence of $k_{-A}$ is weaker than
that of $k_{+A}$.  We therefore neglect the strain-dependence of detachment
rates and assume that the Boltzmann factor in Eq.~(\ref{eq:det_bal}) only
influences $k_{+A(i)}$.  If the binding process is reversible, $P_i$
(Eq.~\ref{eq:bprob1}) becomes independent of this assumption in any case.

To describe the fluctuations in the actin helix, we will use the coefficients
$a_i$, which can assume the values -1, 0 or 1.  This corresponds to the
assumption that each subunit can have three orientations relative to its left
neighbor \cite{Egelman1997}.  The twist between the subunits $i-1$ and $i$ is
then $\theta_{i}-\theta_{i-1}=\theta^0+a_i \theta'$, and the total twist of subunit $i$ relative to subunit $0$ is $\theta_i =i \theta^0 +
(a_1+a_2+\ldots a_i)\theta'$.

Another important question is the dynamics of the actin fluctuations. Let us
denote the probability that the helix is in the state with angles $\theta^0+a_1
\theta'$, $\theta^0+a_2\theta'$, $\ldots$ with
\begin{equation}
  \label{eq:phelix}
  P_{a_1,a_2,\ldots, a_i}= \left(\frac{1}{3}\right)^i \;.
\end{equation}

If the fluctuations are fast in comparison with the attachment rate, the
probability that the head binds to site $i$ is proportional to the attachment
rate in each configuration, weighted by the probability of that configuration
and summed over all possible configurations:
\begin{equation}
  \label{eq:probfast}
  P^f_i=\frac{\sum\limits_{a_1,a_2,\ldots a_i}  P_{a_1,a_2,\ldots, a_i} e^{-U_i(\theta^0 i + \theta'(a_1+a_2+\ldots
  a_i))/k_BT}}
{\sum\limits_j \sum\limits_{a_1,a_2,\ldots a_j}  P_{a_1,a_2,\ldots, a_j}   e^{-U_j(\theta^0 j + \theta'(a_1+a_2+\ldots
  a_j))/k_BT}}
\end{equation}
Here $U_i(\theta)$ denotes the elastic energy in a state where the lead head is
bound $i$ subunits in front of the trail head and the helical twist between
these two subunits is $\theta$.  A simplified expression for $P^f_i$ is derived
in the Appendix.

If, on the other hand, the fluctuations are slow, the probability is given by
the ensemble-average of all helix conformations:
\begin{equation}
  \label{eq:probslow}
  P^s_i=\sum\limits_{a_1,a_2,\ldots}  P_{a_1,a_2,\ldots, a_n} \frac{e^{-U_i(\theta^0 i + \theta'(a_1+a_2+\ldots
  a_i))/k_BT}}
{\sum\limits_j e^{-U_j(\theta^0 j + \theta'(a_1+a_2+\ldots
  a_j))/k_BT}}
\end{equation}
Here $n$ denotes the maximum index of a site that still has a non-negligible
binding probability.  In the following, we will calculate the results for both
scenarios, even though we consider the slowly fluctuating scenario more
realistic. 

\begin{figure}
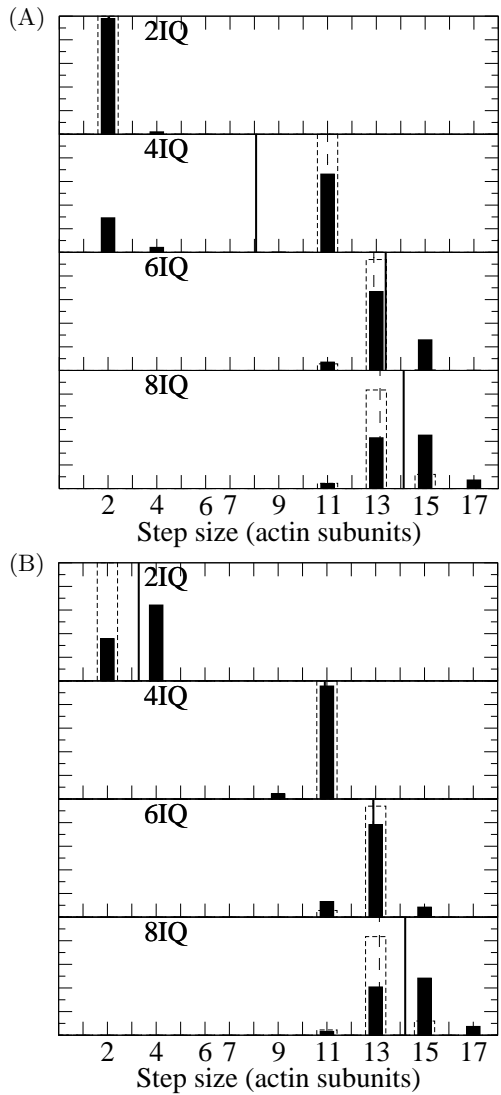

  \begin{center}
\raisebox{7cm}{(A)} \includegraphics[width=6cm]{figure4a.eps}\\
\raisebox{7cm}{(B)} \includegraphics[width=6cm]{figure4b.eps}\\
  \end{center}
  \caption{The probabilities that the lead head binds to site $i$ in the
    model with slow fluctuations ($P_i^s$, A) and fast fluctuations ($P_i^f$,
    B).  The four diagrams are for different lever arm lengths (2IQ: 10nm; 4IQ:
    18nm; 6IQ: 26nm; 8IQ: 34nm).  The vertical lines in each diagram show the
    average position $\left<i\right>$. The dashed lines represent results on a
    rigid 13/6 helix, as used in previous calculations.}
  \label{fig:histo2813}
\end{figure}

A numerical evaluation of the probability distributions $P_i^s$ and $P_i^f$ for
four different lever arm lengths, corresponding to mutants with 2, 4, 6 (wild
type) and 8 IQ motifs in the lever arm shows a notable difference in comparison
with results on the 13/6 helix (Fig.~\ref{fig:histo2813}).

To check the effects of uncertainties in other model parameters on our
results, we have tried three modifications: we have introduced an additional
azimuthal component of the power-stroke (meaning that after Pi release, the
lever arm makes an additional swing of $20^\circ$ to the right, while its
starting point remains the same), we have changed the lever arm angle $\phi$ in
the pre-powerstroke state and we have changed the lever arm stiffness.  The
results are summarized in Table~\ref{tab:robustness}.  None of the tested
modifications makes an improvement with regard to agreement with experimental
data as compared to the basic model.

Another assumption we made in our model is that the lever arm anchoring is
stiff in every state.  While this assumption is justified for long lever arms,
the effect of compliance in the converter domain could play an important role
for short lever arms.  Results on skeletal muscle myosin (with short lever
arms) indeed show that at least half of the compliance is in the converter
domain \cite{Kohler.Kraft2002}.  Therefore some care has to be taken when
interpreting the results for 2IQ lever arms.

\begin{table*}
  \caption{Average step size for the basic model and the following variations:
    (i) additional azimuthal power stroke, (ii) different lever arm angle in
    the pre-powerstroke conformation ($\phi_{ADP.Pi}=135^\circ$ instead of
    $115^\circ$) and (iii) a softer lever-arm ($EI=500\,\rm pN\,nm^2$ instead
    of $1500\,\rm pN\,nm^2$).  All data are based on the slow-fluctuation
    scenario.}  \label{tab:robustness}

\medskip

  \begin{tabular}{lllll}
Lever arm length & basic model & $\Delta \theta=20^\circ$ &
$\phi_{ADP.Pi}=135^\circ$ & $EI=500\,\rm pN\,nm^2$\\
\hline
2IQ & 2.0 & 2.28 & 2.6 & 2.1 \\
4IQ & 8.1 & 4.5 & 11.7 & 7.9 \\
6IQ & 13.4 & 14.1 & 14.0 & 13.9 \\
8IQ & 14.1 & 15.1 & 14.7 & 13.2 \\
\hline
  \end{tabular}
\end{table*}

\section*{Discussion}

The results for the slowly fluctuating scenario show a significantly improved
agreement with available experiments \cite{Walker.Knight2000} than the previous
model, based on a stiff 13/6 helix.  In particular, the observation that the
probability for 15 subunit steps is about twice as high as for 11 subunit
steps is well reproduced.  The average step size of 13.4 actin subunits (on a
helix with a half-pitch of 14) also means that the molecule makes a
left-handed rotation of about $8^\circ$ per step, or $0.2^\circ$ per nm
traveled. This means that a freely walking myosin V makes one revolution
around the actin filament every $1.7\,\rm \mu m$, well consistent with a value
of $2.2\,\rm \mu m$, measured by Ali and co-workers \cite{Ali.Ishiwata2002}.
Note that their interpretation (based on the assumption of a 13/6 helix) is
that the steps represent a mixture of 11 and 13 subunit lengths.  However, on a
28/13 helix, data become consistent with a mixture including 15 subunit
steps, which are even more frequent than 11 subunit steps.

The third experiment our model should be tested against is the dependence of
the step size on the lever arm length.  Purcell and
co-workers \cite{Purcell.Sweeney2002} used optical tweezers to determine the step
size of different dimeric constructs.  For wild-type myosin V (6 IQ domains),
they measured a step size of 35nm.  For 4IQ mutants they measured 24nm or 8.7
actin subunits, and for 1IQ mutants 5nm or 2 subunits. Also in this respect the
current model (predicting 13.4, 8 and 2 subunit steps, respectively) noticeably
improves the agreement with experimental data.  The previous model, based on a
stiff helix, predicted a more abrupt step size change between 2 and 4IQ
mutants.

\section*{Appendix}

In Eq.~(\ref{eq:probfast}) the summation over coefficients $a_1,\ldots,a_i$ can
be carried out while keeping their sum constant ($s_i$). This way we obtain the
probability that the subunit $i$ is oriented at angle $\theta^0 i + \theta'
s_i$:
\begin{eqnarray}
  \label{eq:pfast}
  P_{i,s_i}&=\sum\limits_{a_1,a_2,\ldots,a_i; a_1+a_2+\ldots+a_i=s_i}  P_{a_1,a_2,\ldots,
  a_i} \nonumber\\ &=
\frac 1 {3^i} \sum\limits_{k; |s_i|\le k \le (i+|s_i|)/2} \left(
  \begin{array}{c} i\\k \end{array}\right)   \left( \begin{array}{c} i-k\\k-s \end{array}\right) 
\end{eqnarray}
Then expression (\ref{eq:probfast}) can be simplified to
\begin{equation}
  \label{eq:probfast1}
    P^f_i=\frac{\sum\limits_{s_i=-i}^i P_{i,s_i}  e^{-U_i(\theta^0 i + \theta'
    s_i)/k_BT}}{\sum\limits_j \sum\limits_{s_j=-j}^j P_{j,s_j}  e^{-U_j(\theta^0 j + \theta'
    s_j)/k_BT}} \;.
\end{equation}
For large values of $i$ the central limit theorem can be applied and the
distribution $P_{i,s_i}$ becomes Gaussian.

\section*{Acknowledgment}
This work was supported by the Slovenian Office of Science (Grants
No.~Z1-4509-0106-02 and P0-0524-0106).

\providecommand{\refin}[1]{\\ \textbf{Referenced in:} #1}


\begin{thebibliography}{10}

\bibitem{Reck-Peterson.Mercer2000}
Reck-Peterson,~S.~L.;\ \ Provance,~Jr.,~D.~W.;\ \ Mooseker,~M.~S.;\ \
  Mercer,~J.~A. \textit{Biochim.\ Biophys.\ Acta} \textbf{2000,} \textsl{1496,}
  36-51.

\bibitem{Mehta.Cheney1999}
Mehta,~A.~D.;\ \ Rock,~R.~S.;\ \ Rief,~M.;\ \ Spudich,~J.~A.;\ \
  Mooseker,~M.~S.;\ \ Cheney,~R.~E. \textit{Nature} \textbf{1999,}
  \textsl{400,} 590-593.

\bibitem{Vale2003b}
Vale,~R.~D. \textit{J.~Cell.~Biol.} \textbf{2003,} \textsl{163,} 445-450.

\bibitem{Burgess.Trinick2002}
Burgess,~S.;\ \ Walker,~M.;\ \ Wang,~F.;\ \ Sellers,~J.~R.;\ \ White,~H.~D.;\ \
  Knight,~P.~J.;\ \ Trinick,~J. \textit{J.~Cell.~Biol.} \textbf{2002,}
  \textsl{159,} 983-991.

\bibitem{De_La_Cruz.Ostap2004}
{De La Cruz},~E.~M.;\ \ Ostap,~E.~M. \textit{Curr.~Opin.~Cell Biol.}
  \textbf{2004,} \textsl{16,} 61-67.

\bibitem{Vilfan2005}
Vilfan,~A. \textit{Biophys.~J.} \textbf{2005,} \textsl{88,} 3792-3805.

\bibitem{Walker.Knight2000}
Walker,~M.~L.;\ \ Burgess,~S.~A.;\ \ Sellers,~J.~R.;\ \ Wang,~F.;\ \
  Hammer,~J.~A.;\ \ Trinick,~J.;\ \ Knight,~P.~J. \textit{Nature}
  \textbf{2000,} \textsl{405,} 804-807.

\bibitem{Egelman.DeRosier1982}
Egelman,~E.~H.;\ \ Francis,~N.;\ \ DeRosier,~D.~J. \textit{Nature}
  \textbf{1982,} \textsl{298,} 131-135.

\bibitem{Egelman.DeRosier1992}
Egelman,~E.~H.;\ \ DeRosier,~D.~J. \textit{Biophys.~J.} \textbf{1992,}
  \textsl{63,} 1299-1305.

\bibitem{Egelman1997}
Egelman,~E.~H. \textit{Structure} \textbf{1997,} \textsl{5,} 1135-1137.

\bibitem{Orlova.Egelman2000}
Orlova,~A.;\ \ Egelman,~E.~H. \textit{Biophys.~J.} \textbf{2000,} \textsl{78,}
  2180-2185.

\bibitem{Rief.Spudich2000}
Rief,~M.;\ \ Rock,~R.~S.;\ \ Mehta,~A.~D.;\ \ Mooseker,~M.~S.;\ \
  Cheney,~R.~E.;\ \ Spudich,~J.~A. \textit{Proc.\ Natl.\ Acad.\ Sci.\ USA}
  \textbf{2000,} \textsl{97,} 9482-9486.

\bibitem{Rosenfeld.Sweeney2004}
Rosenfeld,~S.~S.;\ \ Sweeney,~H.~L. \textit{J.~Biol.~Chem.} \textbf{2004,}
  \textsl{279,} 40100-40111.

\bibitem{Snyder.Selvin2005}
Snyder,~G.;\ \ Syed,~S.;\ \ Goldman,~Y.;\ \ Selvin,~P. \textit{Biophys.~J.}
  \textbf{2005,} \textsl{88,} 2450-Pos, Part 2 Suppl. S.

\bibitem{Snyder.Selvin2004}
Snyder,~G.~E.;\ \ Sakamoto,~T.;\ \ Hammer,~J.~A.;\ \ Sellers,~J.~R.;\ \
  Selvin,~P.~R. \textit{Biophys.~J.} \textbf{2004,} \textsl{87,} 1776-1783.

\bibitem{mehta99}
Mehta,~A.~D.;\ \ Rock,~R.~S.;\ \ Rief,~M.;\ \ Spudich,~J.~A.;\ \
  Mooseker,~M.~S.;\ \ Cheney,~R.~E. \textit{Nature} \textbf{1999,}
  \textsl{400,} 590.

\bibitem{Kohler.Kraft2002}
K{\"o}hler,~J.;\ \ Winkler,~G.;\ \ Schulte,~I.;\ \ Scholz,~T.;\ \ McKenna,~W.;\ \
  Brenner,~B.;\ \ Kraft,~T. \textit{Proc.\ Natl.\ Acad.\ Sci.\ USA}
  \textbf{2002,} \textsl{99,} 3557-3562.

\bibitem{Ali.Ishiwata2002}
Ali,~M.~Y.;\ \ Uemura,~S.;\ \ Adachi,~K.;\ \ Itoh,~H.;\ \ {Kinosita Jr},~K.;\ \
  Ishiwata,~S. \textit{Nat.~Struct.~Biol.} \textbf{2002,} \textsl{9,} 464-467.

\bibitem{Purcell.Sweeney2002}
Purcell,~T.~J.;\ \ Morris,~C.;\ \ Spudich,~J.~A.;\ \ Sweeney,~H.~L.
  \textit{Proc.\ Natl.\ Acad.\ Sci.\ USA} \textbf{2002,} \textsl{99,}
  14159-14164.

\end{thebibliography}
\end{document}